\documentclass[graybox]{svmult}

\usepackage{type1cm}        
\usepackage{makeidx}         
\usepackage{graphicx}        
\usepackage{multicol}        
\usepackage[bottom]{footmisc}

\usepackage{newtxtext}       %
\usepackage{newtxmath}       

\def\lya{Ly$\alpha$}

\makeindex             

\begin{document}

\title*{Starburst Galaxies}
\author{Ivana Orlitova} 
\institute{Ivana Orlitova \at Astronomical Institute of the Czech Academy of Sciences \email{orlitova@asu.cas.cz}
}
%
%
\maketitle

\abstract{
The rate of star formation varies between galaxy types and evolves with 
redshift. Most stars in the universe have formed in episodes of an exceptionally
high star-forming activity, commonly called a starburst. 
We here summarize basic definitions and general properties of starbursts, 
together with their observational signatures. We overview the main types of 
starburst galaxies both in the local universe and at high redshift, where 
they were much more common. We specify similarities and differences between
the local and distant samples and specify the possible evolutionary links.  
We describe the role of starburst galaxies in the era of cosmic reionization,
relying on the most recent observational results. 
	}

\section{Star formation and starburst}
\label{sec:SF}


\begin{figure}[hb]
\sidecaption
\includegraphics[width=0.5\textwidth]{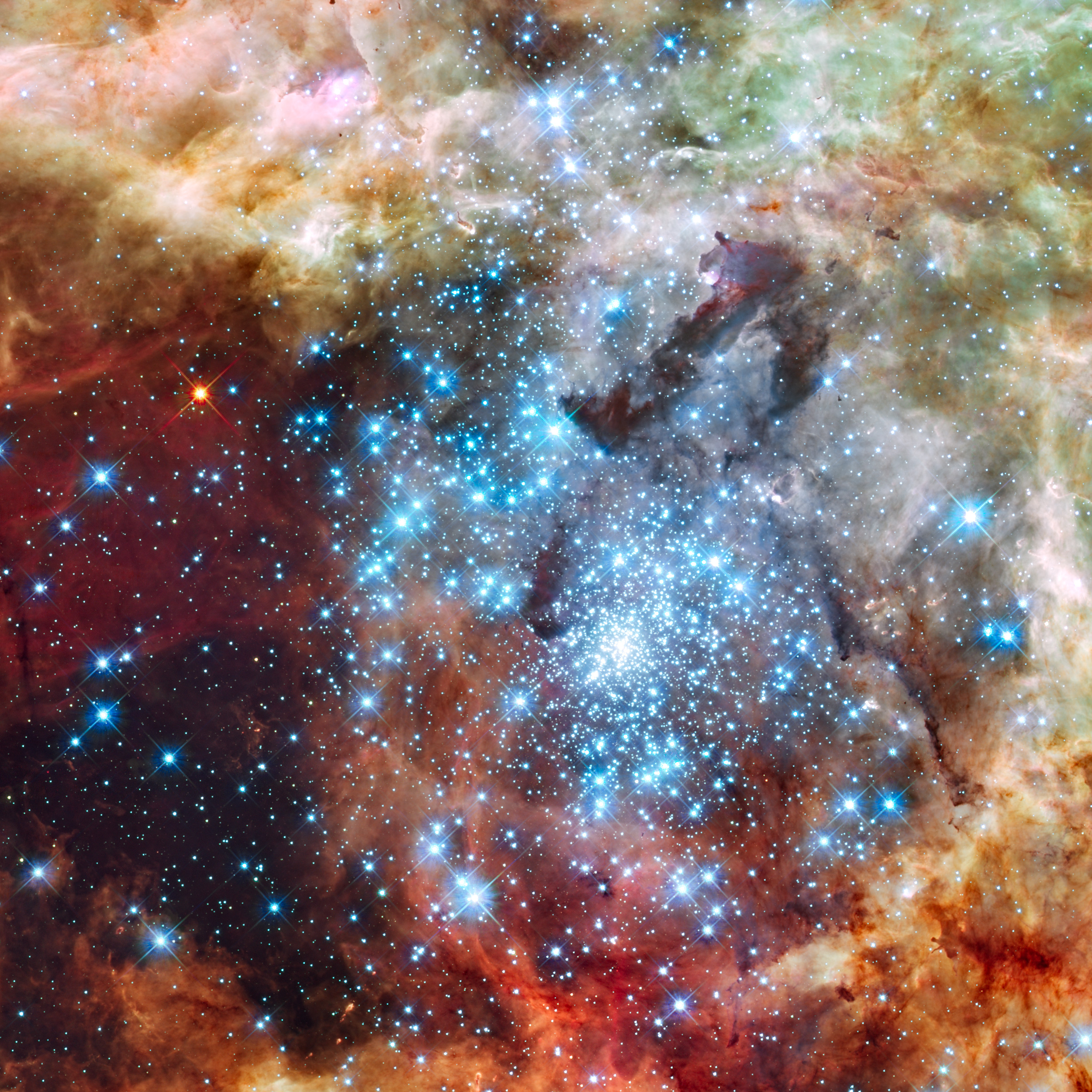}
\caption{Starburst region 30~Doradus (also known as Tarantula nebula) 
in the Large Magellanic Cloud. 
The multi-colour UV and optical image was obtained using the 
Hubble Space Telescope and shows the central concentration of young stars 
(called R136) in the NGC~2070 star cluster, which is situated at the core 
of the nebula. R136 contains several O stars and Wolf-Rayet stars and
produces most of the energy that makes the 30~Doradus visible. 
Credit: NASA/ESA.}
\label{fig:30Dor}
\end{figure}

The rate of star formation is one of the basic characteristics of a galaxy. 
The stellar content and stellar age provide observational signatures that 
allow us to derive the history of galaxy evolution.   
The star formation rate (SFR), defined as dM$_*$/dt, can range from zero, 
such as in gas-poor elliptical galaxies, 
to hundreds or thousands of solar masses per year ($M_\odot$\,yr$^{-1}$) 
in the most vigorously star-forming galaxies. 
For illustration, the Milky Way (MW) forms stars at a rate of 
less than $1\,M_\odot$\,yr$^{-1}$. 
Star formation (SF) can be continuous and more-or-less regular, 
or it can consist 
of short and intense bursts separated by long intervals of quiescence.
Understanding how galaxies make stars and what drives the 
differences between them 
is an active field of research. We will here overview the main observational 
methods and results, focusing specifically on galaxies that undergo a burst 
of star formation. 

The SFR alone may not be a sufficient parameter to 
characterize how important is the SF in a given galaxy. 
In principle, the more massive the galaxy, the larger the SFR
can be, if enough cold gas ($\sim10$\,K) is available.
It is convenient to define relative quantities that  
allow a better comparison between galaxies. 
By factoring out the galaxy stellar mass, we obtain 
the specific star formation rate (sSFR), the SFR per unit mass. 
The sSFR is proportional to the birthrate parameter $b$, the 
ratio of the current SFR to the average past SFR \cite{Kennicutt98}.
Another useful quantity 
is the surface density of star formation, $\Sigma_\mathrm{SFR}$, 
which measures the SFR per unit area and which correlates with the 
gas surface density $\Sigma_\mathrm{gas}$ (combined atomic H\,{\sc i} 
and molecular H$_2$). 
This relation is commonly called `Schmidt-Kennicutt' and has been discussed
in detail in Kennicutt's review~\cite{Kennicutt98}. An updated picture, 
using recent multi-wavelength data, clarifies that the relation is mainly 
driven by molecular $\Sigma_\mathrm{H_2}$~\cite{Bigiel08,Kennicutt12}. 

Star formation proceeds through the assembly of   
cold gas into dense clouds which eventually undergo gravitational collapse
once they attain masses of $\sim10^6\,M_\odot$ and sizes up to 100\,pc. 
What drives this gas compression and what triggers the SF is 
a subject of many discussions as we will present in the following 
sections (shock waves, stellar density waves such as bars, tidal interactions). 
How is the available gas converted into stars is then described 
by the SFR efficiency $\epsilon = SFR/M_\mathrm{gas}$ (definition used
in extragalactic studies, different from that for individual clouds  
in the MW). The inverse of the efficiency is called
the depletion time, which states how long can the galaxy continue forming
stars at the current rate with the given 
mass of atomic H\,{\sc i} and molecular H$_2$. 
A large spread of efficiencies is observed in local galaxies, with an 
average of $\sim\!5\%$ per $10^8$ years~\cite{Kennicutt98,Bigiel08,Kennicutt12},
corresponding to a  
depletion time of 2\,Gyr (which can in reality be a factor of two longer 
due to recycling of gas from stellar winds).

\bigskip

When a galaxy undergoes an exceptionally intense phase of SF, we 
speak of a starburst (Figure~\ref{fig:30Dor}). 
A starburst galaxy is thus not a separate class and 
it is rather an evolutionary phase in the life of a galaxy. 
The starburst activity may not be uniformly spread across the galaxy. 
While global starbursts preferentially occur in dwarfs,  
the starbursts in massive galaxies 
are usually localized in a small volume,  
most often in the circumnuclear region, where 
$10^8-10^{10}$\,M$_\odot$ 
of gas is confined to a radius of  $<\!1$\,kpc.  
The reasons behind the vigorous star formation are not yet fully understood. 
The most appealing is the effect of 
gravitational interaction 
between galaxies, ranging from a close passage to a complete merger
(from pioneering papers such as \cite{Larson78,Duc94} to recent 
observations~\cite{Ellison08} and simulations~\cite{DiMatteo07}).
However, it is not yet clear whether such an encounter 
is a necessary condition 
(some starburst galaxies seem to be isolated), 
nor whether the encounter is a sufficient condition (tidal forces  
between dwarf galaxies are modest).  
The tidal forces perturb the orbits of gas and stars and enable the gas flow 
toward the galaxy center. A similar transfer of gas is also possible by 
the action of gravitational instabilities such as stellar 
bars~\cite{Hopkins10,Elmegreen12}. 
Local dynamical processes including the pressure, the local velocities, 
the starlight and 
self-gravitation decide on the future of the gas clouds~
(as reviewed by \cite{Elmegreen12,Elmegreen14}).
The accumulation of gas in the central parts is essential 
for yet another phenomenon:
feeding of the active galactic nucleus (AGN), i.e. 
the central galactic region dominated by a supermassive black hole. 
Indeed, a simultaneous AGN and starburst activity  
exists in numerous galaxies, as we will describe in the following sections.  
It is still a matter of debate if the coexistence is coincidental or if 
there is a causal relationship ~\cite{Alexander12}.  
If no apparent causes of starburst are visible,  
we can speculate about the time delays between different phenomena or invoke 
the role of shock waves due to preceding star formation and supernova 
explosions~\cite{Gerola78,Elmegreen12}.

\bigskip

There is no unique and rigorous definition of a starburst, 
and a good definition is being discussed in the literature still 
today~\cite{Heckman05b,Kennicutt12,Bergvall11}. 
The classical definition is related to
the amount of gas in the galaxy and the maximum lifetime of the starburst: 
the galaxy would consume all of its 
remaining gas in a limited time were the SFR constant. 
If this consumption (depletion) time is much shorter than the age of 
the universe, then we define the galaxy as a starburst. 
In other words, 
the efficiency of SF must be at least an order higher than 
in an average local galaxy~\cite{Kennicutt98}.
However, the problem with this definition is the uneven gas 
fraction in different 
galaxy types. The depletion time in a massive, gas-poor galaxy can be short, 
yet without any sensible reason to call it a starburst.    
An alternative definition is based on the birthrate parameter:
one can estimate the time that the galaxy would need in order to
create all of its existing stars at the present SFR. A disadvantage in this 
case is for the massive galaxies where the 
timescales to reproduce their stellar mass are excessively long. 
Moreover, both of these approaches have a built-in redshift dependence -- the 
varying age of the universe changes the quantitative meaning of 
starburst parameters at each redshift~\cite{Heckman05b}.
One therefore has to be careful what is meant by a starburst in 
each study. Heckman~\cite{Heckman05b} proposed  
to define a starburst based on $\Sigma_\mathrm{SFR}$,
which treats equivalently both nearby and high-redshift galaxies. 
High $\Sigma_\mathrm{SFR}$ implies 
large surface densities of both gas and stellar mass \cite{Kennicutt98}. 
As a consequence, such regions will 
have a high gas pressure, a high rate of mechanical energy deposition 
and a high density of radiant energy. $\Sigma_\mathrm{SFR}$ 
is hence directly linked to the galaxy physical properties.

The definition of a starburst may have to stay loose, we can nevertheless 
require a parallel application of several criteria, and, in addition, 
add a qualitative requirement that the starburst must have a  
global impact on its host galaxy evolution -- powerful outflows,  
significant mass fraction transformed to stars, disturbed structure, 
enhanced luminosity, etc. 
Typically, authors require $b\geq3$ or $b\geq10$ to select starburst 
galaxies~\cite{Ostlin01,Bergvall16}. 
The undisputed starbursts have durations of $10^7 - 10^8$\,years and  
$\Sigma_\mathrm{SFR}\sim 1-100$\,M$_\odot$\,yr$^{-1}$\,kpc$^{-2}$, 
which exceeds $\Sigma_\mathrm{SFR}$ of the MW by several orders of 
magnitude~\cite{Heckman05b}. 
The most extreme of them form stars with efficiencies close to 100\% per
$10^8$\,yr~\cite{Kennicutt98}.  
What is certain is that the upper bound for efficiency must be set by 
causality -- conversion 
of the entire self-gravitating gas reservoir into stars in one dynamical time.

Starbursts (defined by $b\geq3$) currently form 20\% of the present-day 
massive stars \cite{Heckman05b}, representing a non-negligible 
baryonic constituent 
of the local universe (if $b\geq10$ is considered, the fraction drops to 3\%). 
Because of their young age, they contain large numbers 
of massive stars and X-ray binaries, and they thus offer unique 
opportunities for studying the high-mass objects and their feedback in the form
of mechanical energy (winds, jets) and energetic photons (ultraviolet, X-rays).
The collective action of stellar winds, supernovae and jets drives powerful 
outflows of gas on the galactic scale. Also known under the name superwinds, 
the large-scale outflows transport material, including heavy elements, 
to the intergalactic medium (IGM). The localized starburst hence directly 
affects a large volume of space.   
Starburst galaxies were much more frequent in the early universe than they are 
today and they played an important role in shaping the IGM
throughout the history of the universe. 
Besides, they were responsible for forming most of the stars in the universe
as they were the building blocks of present-day galaxies.
SF then sharply dropped after redshift $z\sim2$ (i.e. 
age of the universe $\sim3$\,Gyr). 
The first starbursts also played yet another role 
in cosmology: the ultraviolet radiation of young massive stars was probably 
responsible for converting the IGM from neutral to ionized in the first
billion years after the Big Bang (so called Cosmic Reionization). 

\bigskip
The goal of this review is to provide a basic description of how are 
starbursts observed and how the observational methods lead to their 
``classification'' both at low and high redshift. 
Section~\ref{sec:obs} describes the starburst signatures at various spectral 
wavelengths.  
Section~\ref{sec:local} overviews the major starburst galaxy types in the 
local universe. 
Section~\ref{sec:highz} presents the high-redshift galaxy 
classes, all detected thanks to the starburst signatures. 
Section~\ref{sec:cosmic} focuses on the role of starburst galaxies in the
cosmic reionization. 
Finally, Section~\ref{sec:concl} brings concluding remarks and future 
prospects, 
including the observational facilities that are in preparation.

\section{Observational signatures of a starburst}
\label{sec:obs}

Star formation manifests itself by a variety of features across 
a large portion of the electromagnetic spectrum. We observe the 
direct starlight of young stars as well as the light reprocessed 
by interstellar gas and dust.  
Detailed and insightful reviews of SFR indicators 
were provided by Kennicutt \cite{Kennicutt12} and Calzetti \cite{Calzetti12}.

\subsection{Ultraviolet and optical continuum}
\label{sec:UV}

The UV and optical light of young stellar populations is dominated by 
massive stars with the lifetimes of $10^6-10^8$ years. 
The O and B stars produce most of their energy in the far-ultraviolet (FUV)
band, including a large fraction 
at Lyman continuum wavelengths ($\lambda<912$\,\AA),
capable of ionizing hydrogen atoms. 
In parallel, massive stars drive powerful stellar
winds that affect kinematics of the interstellar medium (ISM) 
and produce specific spectral features. 
The importance of the UV observations in starbursts was demonstrated by the 
International Ultraviolet Explorer (IUE) satellite, operating for almost 
twenty years
before the end of the 20th century. The IUE was followed by the Far Ultraviolet 
Spectroscopic Explorer
(FUSE), the Galaxy Evolution Explorer (GALEX), and the Hubble Space 
Telescope (HST). 
Thanks to the IUE, the first UV catalogue of stellar types 
was issued in 1985~\cite{Walborn85}.  

In the dust-poor case, the bright UV continuum is a prominent signature  
of the young massive stars, i.e. of a starburst.  
The bright UV is also observable for high-redshift galaxies
and represents one of the major methods of galaxy detection   
(Section~\ref{sec:LBGs}, Figure~\ref{fig:LBG}). 
Furthermore, empirical relations exist between the SFR and the UV luminosity, 
and therefore 
both FUV and near-UV (NUV) luminosities serve as SFR estimators
\cite{Calzetti12}. 
Nevertheless, the total SFR estimation must ideally take into account 
the effect of dust. The dust efficiently reprocesses the UV light 
into infrared (IR) radiation (see Section~\ref{sec:IR}). 
Therefore, combined UV and IR measurements
provide a more precise determination of the galaxy SFR~\cite{Calzetti12}.

The FUV light short of the Lyman edge ($\lambda=912$\,\AA) 
ionizes neutral hydrogen. As a result, most of this 
radiation is absorbed in the ISM or IGM and does not 
reach our telescopes.  
Instead, a sharp drop, known as the Lyman break, appears in the FUV spectrum. 
This property is conveniently used for determining the galaxy redshift at 
various epochs of cosmic history
(Section~\ref{sec:LBGs}, Figure~\ref{fig:LBG}). 
Part of the absorbed ionizing radiation is reprocessed into emission 
lines which  
are formed by recombination and which become another SF signature  
(Sections~\ref{sec:Lya}~and~\ref{sec:lines}).
If, on the other hand, the column densities of interstellar 
H\,{\sc i} are low enough 
not to remove all of the Lyman continuum, the remaining radiation will escape 
to the surrounding IGM and ionize it 
-- which is of great interest mainly to the high-$z$ studies and the 
era of cosmic reionization ($z>6$, Section~\ref{sec:LyC}).

\subsection{UV absorption lines}

\begin{figure}[hb]
\includegraphics[width=1.\textwidth]{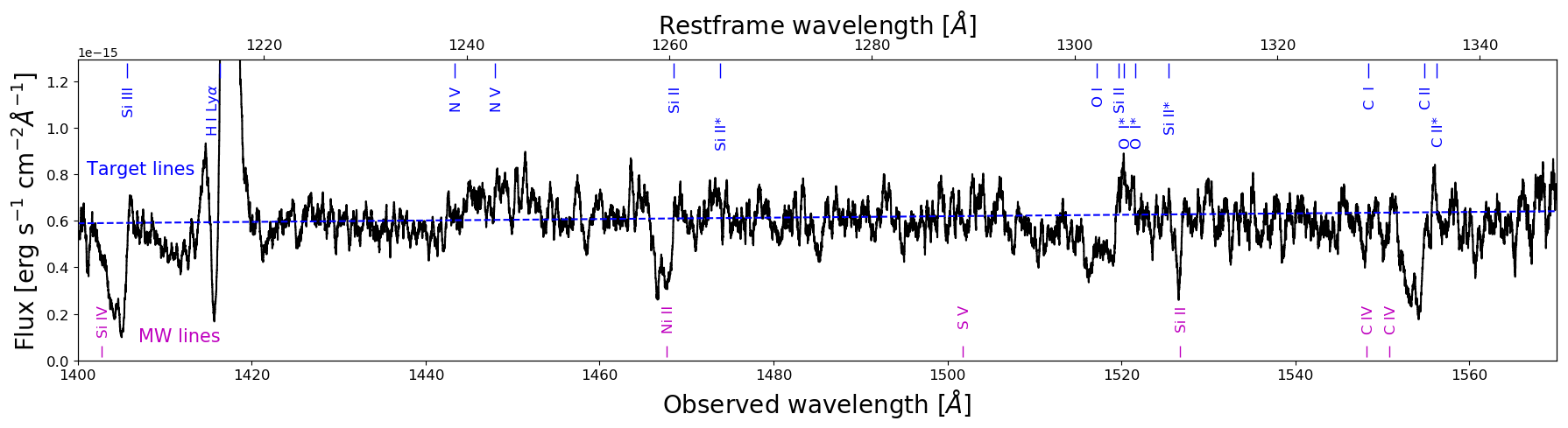}
\caption{UV absorption lines of the galaxy SDSS-J030321.41-075923.2. 
The major lines originating both in the galaxy and in the Milky Way have been
labeled. 
Source: HST archive.}
\label{fig:UVlines}
\end{figure}

UV-bright galaxies allow observation of starburst properties through 
UV absorption lines formed in the foreground ISM (Figure~\ref{fig:UVlines}). 
The lines are especially well
detectable in starbursts producing copious amounts of young stars and 
thus UV photons. 
The lines inform us about chemical composition of the ISM gas and its 
ionization state (e.g. Si\,{\sc ii}, Si\,{\sc iii}, 
Si\,{\sc iv}), about kinematics and about optical depth of the gas. 
The FUSE satellite opened the way to studying the UV lines in the MW and 
nearby galaxies, as it was the first 
UV facility with sufficient spectral resolution. 
The HST has then revolutionized the
field with progressively more sensitive spectrographs and improving resolution: 
GHRS, STIS, COS. 
Conversely, at high redshift, UV lines are reachable with ground-based 
optical telescopes, which need to have a sufficient collecting area and 
a spectrograph of sufficient resolution. 

The UV absorption lines probe gas kinematics in nearby starbursts, 
showing that their properties may 
range from static to high-velocity ($\sim\!1000$\,km\,s$^{-1}$) 
outflows~\cite{Heckman11,Chisholm15}, and may vary between different gas phases.
Outflows were also found to be ubiquitous at high redshift~\cite{Shapley03}. 
The UV absorption lines thus offer useful tools for answering questions 
about stellar feedback, ISM enrichment, IGM enrichment, gas flows, 
and galaxy evolution.     
We describe in Section~\ref{sec:Lya} that the UV lines also represent  
valuable complementary information for interpreting the Lyman-alpha line. 
Eventually, the optical depth of the UV lines 
has been explored as a tool for studying
the escape of ionizing UV continuum from galaxies~\cite{Heckman11}.
If the neutral gas situated along the line of sight does not completely 
cover the ionizing source, we should observe residual flux in the 
absorption lines. This can in turn be translated to the Lyman continuum 
escape fraction~\cite{Savage91}. 
However, in reality, the interpretation of residual fluxes is not 
straightforward, is dependent on the gas geometry, its chemistry, on spectral 
resolution, and on the applied model. 
Observational testing of the method against the directly measured 
Lyman continuum escape has only started recently, 
after the first successful Lyman continuum detections have been 
confirmed~\cite{Chisholm17,Gazagnes18}.

\subsection{Lyman-alpha line of hydrogen (UV)}
\label{sec:Lya}

%
\begin{figure}[b]
\sidecaption
\includegraphics[scale=.45]{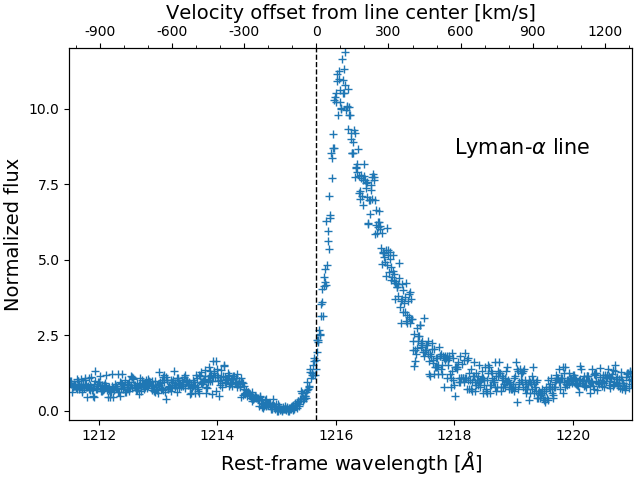}
%
%
\caption{
\lya\ spectrum of Mrk~259, 
a nearby irregular starburst galaxy from the 
LARS sample. The P-Cygni profile with its asymmetric red peak and 
the absorption trough blueward from the line center 
(1215.67\,\AA, dashed vertical line) 
result from radiative transfer. 
Credit: HST archive. 
}
\label{fig:Lya}       
\end{figure}

Lyman-alpha (Ly$\alpha$), at wavelength $\lambda=1215.6$\,\AA, 
is the first transition of hydrogen and in principle 
its brightest spectral line. In galaxies, it is produced by the recombination 
process in the regions of ionized gas mostly around hot stars and AGN.
The probability that the recombination cascade passes through the first 
transition is $\sim70\%$ and  
Ly$\alpha$ thus reprocesses a major part of the stellar ionizing radiation. 
It may hence appear as the ideal signature of star-forming galaxies 
and it was 
indeed predicted to be the beacon of high-$z$ galaxies as early 
as the 1960s~\cite{Partridge67}.
However, the reality is less straightforward: \lya\ undergoes 
a complex radiative transfer in the neutral ISM of the galaxy. By symmetry, 
its 
cross-section for interaction with neutral hydrogen is large and the line 
becomes optically thick at H\,{\sc i} column densities as low as 
$10^{13}$\,cm$^{-2},$ i.e. basically in all galaxies 
(see a useful review of \lya\ physics by M.\,Dijkstra \cite{Dijkstra14}). 
Observational confirmation of \lya\ only came almost 
twenty years after the prediction and the line was far weaker than 
expected (\cite{Meier81} in low redshift; \cite{Djorgovski85} in $z\sim3$;
first HST \lya\ images in~\cite{Kunth03}).

The \lya\ physics started to emerge with the growing observational samples   
both at low and high redshift, and, in parallel, with the development of 
numerical models. Building on analytical solutions that were only 
possible for a few extreme cases~\cite{Neufeld90}, the numerical codes 
explore a range of geometries and conditions, from 
homogeneous spherical set-ups~\cite{Dijkstra06,Verhamme06},
through clumpy media~\cite{Gronke17} to full hydrodynamic simulations
\cite{Verhamme12}. 
The resonant scattering
off hydrogen atoms increases the probability of \lya\ destruction by 
dust, therefore the dust role is enhanced with respect to other UV wavelengths.
The \lya\ line is further sensitive to the geometry of the
medium, its clumpiness, and the macroscopic and microscopic kinematics. 
To escape from the ISM, the \lya\ photons must either shift to 
regions where the H\,{\sc i} density is low, 
or shift in frequency to the line 
 wings (by interaction with hydrogen atoms) 
 where the scattering cross-section is smaller. 
As a result, \lya\ galaxy images commonly show a ``halo'', which here means
a low surface brightness \lya\ emission extending beyond the stellar continuum 
image~\cite{Hayes13,Momose14,Wisotzki16}. This feature is likely due to 
photons produced in the starburst regions and transferred to the outskirts 
of the galaxy.
Conversely, the \lya\ escape through a frequency shift is detected 
in spectral line profiles, which range from  
asymmetric P-Cygni 
(absorption in the blue, emission in the red, see Figure~\ref{fig:Lya}), 
through single red peaks (mostly in high $z$), 
double- or multiple-peaks, 
to damped absorption profiles (with no 
escaping photons). 
All of the profiles result from radiative transfer of photons 
from the line core to the line wings. 

Despite its complex interpretation, \lya\ is one of the primary tools
for galaxy detection at high redshift (Section~\ref{sec:LAEs}). 
Now that we are aware of its limitations and we search in the correct 
luminosity range (much fainter than the theoretical one), \lya\ stays 
detectable out to redshifts beyond $z\sim10$. As demonstrated in the local 
universe (from the earliest studies such as~\cite{Meier81,Kunth98}), 
the \lya\ emission is only present in a subset of star-forming galaxies and 
therefore introduces selection biases that may not be fully understood 
today. 
\lya\ is mostly bright at low-mass, low-metallicity, dust-poor 
starburst galaxies (not speaking about quasars here). 
However, due to the multi-parameter nature of \lya\ transfer, 
we do not have a full control of the galaxy populations detected in \lya,
and complementary methods are necessary for characterizing 
galaxy evolution at high redshift.  
A good lesson was provided by the extremely low-metallicity, 
dust-poor local galaxy IZw18, where bright \lya\ emission was expected 
and where the HST revealed a deep absorption instead~\cite{Kunth94}. 
We understand now from the UV metal lines that the neutral gas is static 
in IZw18, which disables the \lya\ escape. We have to take into account 
these effects when interpreting high-$z$ samples, especially 
in experiments where only emission is targeted.  

At present, \lya\ imaging and spectroscopy
are among the main objectives of all currently developed facilities 
such as the James Webb Space Telescope (JWST) 
and the generation of extremely large telescopes (ELTs).  
The Subaru telescope has its powerful Hyper-Suprime Cam, the ESO's 
Very Large Telescope (VLT) recently obtained the highly performant 
integral-field spectrograph MUSE optimized for \lya, its successor 
HARMONI is in development for the ESO's ELT.
The Hobby Eberly Dark Energy Experiment (HETDEX) is expected to detect
millions of \lya\ galaxies. \lya\ will therefore stay a substantial 
cosmological tool in the years to come, despite all its limitations.

\subsection{Optical emission lines}
\label{sec:lines}

Starburst galaxies are characterized by bright optical emission lines that 
are formed in the ionized interstellar gas (see Figure~\ref{fig:GP}). 
Part of the lines arise from recombination, and represent thus reprocessed 
ionizing radiation of O and B stars. 
This is the case of the hydrogen Balmer series, for instance. 
The H$\alpha$ line can 
reach equivalent widths over thousand \AA\ in starburst galaxies. 
The H$\alpha$ luminosity is also one of the favourite tools for SFR 
determination~\cite{Kennicutt98}.   

Forbidden lines represent another type of transitions, 
excited by collisions and de-excited radiatively. 
Their luminosities are not directly coupled to the ionizing luminosity, 
but they are bright in starburst galaxies 
thanks to the high temperatures and the energy pumped 
into the ISM by star formation. 
Forbidden lines of S$^+$, N$^+$, O$^+$, O$^{++}$ are the most prominent ones.  
Their formation is sensitive to  
metallicity, electron temperature and density, therefore the 
ratios of the line fluxes are useful for measuring the fundamental gas 
properties. In addition, their combinations allow classifying the galaxies to 
AGN and star-forming galaxies.
Statistical classification trends emerged 
from the pioneering work of Baldwin, Phillips and Terlevich 
(their so called BPT diagram~\cite{Baldwin81}) and similar. 
Building on these early works,   
the big data brought about by the Sloan Digital Sky Survey (SDSS) 
at the turn of this century led to the discovery 
of a clear separation between star-forming galaxies and 
AGN into two narrow sequences in the BPT diagram~\cite{Kauffmann03}. 
Theoretical works such as \cite{Kewley01,Stasinska06,Kewley13} 
provided interpretation and prescription for the observational  
patterns. 
The oxygen [O\,{\sc iii}]$\lambda5007$ line is 
among the brightest optical lines under certain conditions. 
Sensitive to the temperature, the line is bright in low-metallicity 
environments (around 10\% solar, e.g. in dwarf starburst galaxies) and in 
high-excitation regions (in AGN).
A second line, such as [N\,{\sc ii}]\,$\lambda6583$ in the BPT diagram,  is
necessary for discriminating between the low-metallicity starburst and 
the AGN. In extreme starbursts, such as those at high redshift, 
the division between SF and AGN becomes 
more delicate: the extreme ISM conditions produce emission line fluxes
originally believed to exist only in AGN. Recent detections of powerful 
starbursts thus make us revise existing models and our understanding of 
star formation~\cite{Kewley13}. 

The optical band also hosts spectral lines that are related to specific
astrophysical questions connected to the starbursts. 
One of the examples is the lines produced by Wolf-Rayet (WR) 
stars, the evolved descendants of O stars. 
A blue bump around 4600\,\r{A} and a red bump around 5700\,\r{A}, 
both formed by a gathering of several emission lines, are characteristic WR 
features. He\,{\sc ii} lines such as 4686\,\r{A} can 
also be of WR origin. The WR evolution is strongly 
affected by metallicity: at low metallicities, only the most massive stars 
become Wolf-Rayet, probing thus the high-mass end of the initial mass function 
(IMF). For more details on WR observations, we refer to~\cite{Kunth00}. 
Another example is the use of optical lines for probing the primordial helium 
abundance. While the oxygen abundance (O/H) varies among galaxies, the helium 
abundance (Y) stays nearly constant. By extrapolating the measured trend to 
zero O/H, one should retrieve the primordial Y. Observations of 
low-metallicity star-forming dwarfs play a decisive role in this 
problem~\cite{Kunth00}.

Detection of the brightest rest-frame optical lines 
(H$\alpha$, H$\beta$, [O\,{\sc iii}], [O\,{\sc ii}]) 
in high redshift 
has been achieved in the last decade using the 8-10\,m 
telescopes such as the Very Large Telescope (VLT), 
the Keck Telescope or Subaru. 
Future larger samples and a more complete coverage of lines 
will provide important clues to the conditions in the distant galaxies.

\subsection{Infrared emission}
\label{sec:IR}

Star formation takes place in dense molecular and dusty clouds.   
It has been approximated that dust absorbs half of the stellar light produced  
in the universe and re-emits it in the infrared. As the absorption 
is particularly efficient in the UV, the IR band is an important complement 
to the UV SFR estimators. The IR alone is not sufficient 
to correctly retrieve the SFR, but the combined UV and IR measurements
proved to be an efficient method. The IR correction can be essentially 
negligible in dust-poor dwarf galaxies and metal-poor regions, but its 
importance increases with growing metallicity and dust content.  
On the other hand, the IR emission can 
significantly overestimate the SFR in the cases where evolved stars contribute
to dust heating. 
The conversion factor between dust emission and SFR 
must therefore be a function of the stellar population~\cite{Kennicutt12}.

The far-IR emission is from 99\% composed of continuum radiation 
produced by dust grains. The remainder is line emission from atomic and 
molecular transitions in the ISM gas, concentrated in a small, sub-kpc region.
The IR continuum depends on the dust content and composition, but also on the 
spectrum of the starlight: luminous young stars will heat dust to higher 
temperatures than old stars and their thermal IR 
will thus peak at shorter wavelengths ($\lesssim\!60\,\mu$m). 
The complexity of dust emission 
features corresponding to small and large dust grains at different
temperatures have been summarized in excellent review papers by 
Calzetti~\cite{Calzetti12} and Kennicutt~\cite{Kennicutt12},
together with the IR SFR calibrators. For the purposes 
of this text, we only need the qualitative statement that the IR emission 
provides an important 
window for detecting obscured starbursts at both low and high 
redshift (Sections~\ref{sec:LIRG} and 
\ref{sec:SMGs}).

\subsection{X-rays}

X-rays trace the star-forming activity through its final products -- 
X-ray binaries -- and through accompanying phenomena 
such as heating of the surrounding ISM by stellar feedback and supernovae.  
Binary systems composed of a main-sequence star and 
a compact object -- black hole or neutron star -- 
emit X-rays by accretion of matter,  
transferred from the donor star onto the compact object 
(see review~\cite{Remillard06}). 
The systems are classified as high-mass or low-mass X-ray binaries  
according to the mass of the donor star. 
Starbursts are characterized by prominent X-ray emission that is dominantly 
provided by the rapidly evolving high-mass X-ray binaries (HMXBs).
Additional X-ray luminosity is provided by interstellar hot gas peaking 
at $\sim\!1$\,keV and  a few additional sources such 
as low-mass X-ray binaries or cataclysmic variables. 
In addition, more exotic sources may contribute decisively: ultra-luminous
X-ray sources (ULX) whose nature is debated (extreme X-ray binaries, 
anisotropic emission, neutron stars)~\cite{Kaaret17}, 
and intermediate-mass black 
holes (IMBH). Such objects seem to be more probable at low metallicity and at 
high SFR~\cite{Greene19}. 

X-ray emission has been mapped in a number of local dwarf galaxies 
including the low-metallicity IZw18, SBS0335-052~\cite{Thuan04},  
Haro\,11~\cite{Grimes07} and other 
starbursts \cite{Oti-Floranes12,Oti-Floranes14,Basu-Zych13,Brorby17,Svoboda19}.
Their X-ray luminosity is generally dominated by several extremely bright, 
point-like sources of debatable origin -- usually too bright for
``classical'' X-ray binaries, they are suspected to be ULXs, IMBHs or even AGN
(though often without optical counterparts). 
Empirically, the total X-ray luminosity of star-forming galaxies 
is proportional to the SFR \cite{Ranalli03} and inversely correlates  
with metallicity~\cite{Basu-Zych13,Douna15,Brorby16}. 
The theoretical basis for starburst X-ray emission was provided  
by numerical simulations of X-ray binaries and of hot gas emission, 
in comparison with observational data~\cite{Cervino02,Mas-Hesse08,Mineo14}.
X-ray observations remain challenging at high redshift, where they are  
restricted to images of stacked samples~\cite{Basu-Zych13}. 
Together with sensitive, 
detailed, resolved mapping of nearby sources, the high-$z$ observations 
will be a task for the new generation of X-ray satellites.

\section{Local starburst galaxies} 
\label{sec:local}

We will focus on global starbursts, rather than individual starbursting
regions of massive galaxies 
(such as the central region of M82, or 30~Doradus in the Large Magellanic 
Cloud -- Figure~\ref{fig:30Dor}). The latter are frequent in the local 
universe, the former are rare. 
Global starbursts are interesting from the cosmological 
point of view, as 
they are traceable out to the earliest epochs of the universe, 
where they were bright and numerous. 
The focus on global starbursts here predefines the focus on 
dwarf galaxies and on extremely dusty IR galaxies. 

While dwarfs constitute the most common
galaxy type in the nearby universe ($>\!70\%$), only a few percent of them 
are starbursting. They have attracted attention for many 
decades~\cite{Zwicky65,Sargent70}. Their subsets appeared under 
various names in the past, such as 
H\,{\sc ii} galaxies~\cite{Terlevich91} or blue amorphous 
galaxies~\cite{Gallagher87}, 
reflecting the techniques of their discovery and the focus on a particular 
feature at a time. Some of the starbursting dwarf classes 
will therefore partially or totally overlap. 
Our intention here is to mention those that have been extensively studied 
in recent past with the motivation to search for 
analogies with high redshift.

\subsection{Blue compact dwarf galaxies}
\label{sec:BGGs}

%
\begin{figure}[b]
\sidecaption
\includegraphics[scale=.3]{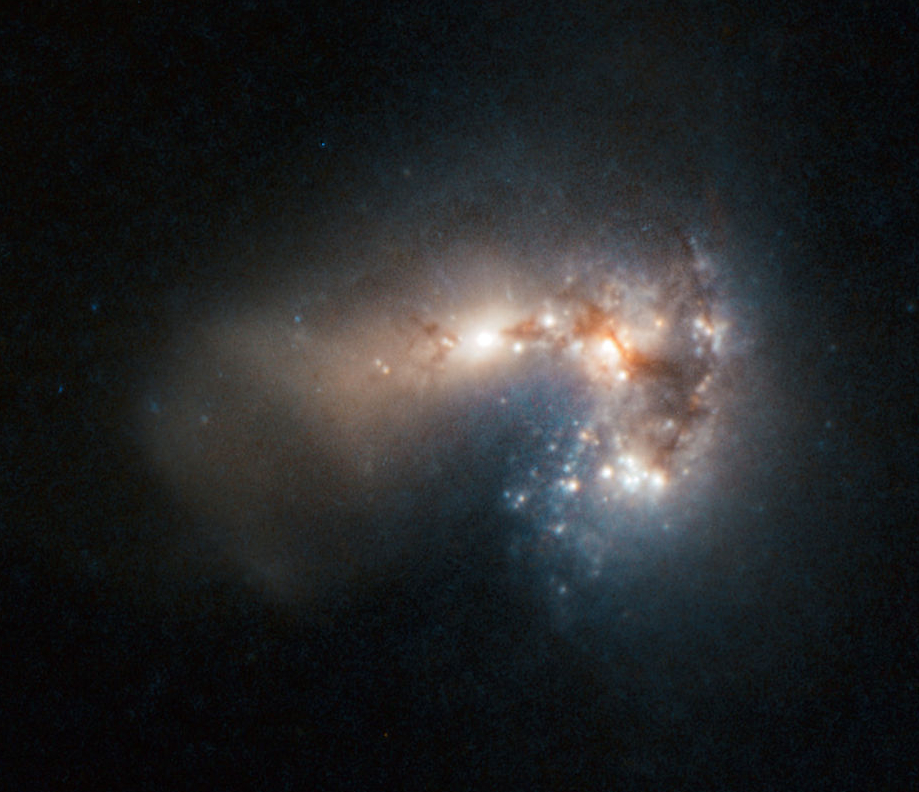}
%
%
\caption{
A multi-color HST image of Haro~11, 
one of the nearest BCGs (and Lyman break analogues). 
Its irregular structure, of total diameter $\sim\!2$\,kpc, 
is probably the result of a galaxy merger, which also 
induced the SFR\,$\sim\!20\,M_\odot$\,yr$^{-1}$. 
The galaxy is a weak Lyman continuum leaker (Section~\ref{sec:LyC}). 
It is a \lya\ emitter in some of the star-forming knots and a 
\lya\ absorber elsewhere.  
Credit: ESO/ESA/Hubble and NASA
(https://www.eso.org/public/images/).
}
\label{fig:Haro11}       
\end{figure}

Blue galaxies, associated with the bright continuum of young stars, 
attracted attention since the 1960s. 
The colour selection alone resulted in a mixed bag of galaxy morphologies and
luminosities. A subset that appeared barely resolved on optical plates of 
that time got the name blue compact galaxies (BCGs). It is still a heterogeneous
class which some authors, with the presently improved knowledge, 
further sub-divide by compactness, luminosity 
or morphology. Not all of the BCGs would comply with all the definitions of 
a starburst.  Their common properties are relatively small mass 
($<10^9M_\odot$), low metallicity, bright optical emission lines, and 
sizes within a few kpc. Galaxies discovered later (for instance by 
emission lines) and having similar properties have been added to the class.  
The most centrally peaked ones ($<2$\,kpc) are referred to as blue compact
dwarfs. 
 
BCGs include some of the most metal-poor galaxies (e.g. IZw\,18, 
see~\cite{Searle72}).  This led
to the thinking that they formed only recently (which would have profound 
implications for galaxy formation). However, this was 
disproved by the discovery of older stellar 
populations (see review~\cite{Kunth00}). 
Instead, their star formation 
has been inefficient for most of the history, interleaved with
time-limited bursts. BCGs are gas-rich ($\gtrsim50\%$ of mass is in gas)
and dust-poor. Their morphologies range from H\,{\sc ii}-region-like 
(Pox\,186) to irregular and clumpy (see Haro\,11 in Figure~\ref{fig:Haro11})
to more symmetric ones. 
Their morphologies and velocity fields often bear signs of a recent 
interaction, despite the relative isolation of many of them  
\cite{Kunth00,Ostlin01,Bergvall11}. 
The presence of low surface brightness neighbours cannot be excluded in such
cases. 
Interaction with neighbours may, in turn, be responsible 
for the starburst activity in BCGs. This comes with a limitation 
that it is uncertain if interaction 
of two dwarfs is sufficient for triggering SF as in 
massive galaxies -- in dwarfs, the gravitational effects are equally 
strong as the 
turbulent velocities caused by stellar winds. 
Starbursts in BCGs challenge the models of star formation anyhow: dwarfs
are generally inefficient in converting gas to stars due to the weak  
gravitational bounding. In addition, the existence of molecular gas 
in BCGs is difficult to 
prove due to their low metallicity. Nevertheless, starbursts in BCGs do exist.

There have been long-standing debates about the evolutionary 
relation between BCGs and other dwarf galaxy classes~\cite{Bergvall11}. 
Answering these questions is 
obviously interesting with respect to the high-$z$ universe, where
galaxies of similar masses were prevailing.   
For the same reason, BCG observations in the UV and \lya\ are important. 
For their low dust amount, a bright \lya\ emission was naively expected in all 
BCGs. This hypothesis was disproved by observing a similar 
incidence of emission and absorption \lya\ profiles~\cite{Kunth98}. 
The extremely metal-poor IZw\,18 is one of the surprising examples of a
deep absorption. 
This finding illustrates the importance of H\,{\sc i} kinematics 
for the \lya\ escape~\cite{Hayes15}, which we described in 
Section~\ref{sec:Lya}.  
The absence of \lya\ emission in a subset of BCGs is an 
important realization for high-$z$ surveys, where the \lya\  
technique will certainly be missing such galaxy populations.

\subsection{Lyman-alpha reference sample}
\label{sec:LARS}

Efforts to understand high-redshift galaxies have led to a definition 
of the local Lyman-alpha reference sample (LARS)~\cite{Ostlin14}.  
The \lya\ line becoming one of the dominant tools for galaxy detection 
at high redshift, the need was to understand the mechanisms of its 
formation and transfer through the ISM. This is achievable in the 
local universe, where spatial resolution and multi-wavelength data 
are available. The complex \lya\ physics (Section~\ref{sec:Lya}) 
does not allow straightforward interpretation of galaxy properties from 
unresolved data, and therefore support from other spectral bands and spatial 
resolution, combined with numerical models is the only way to achieving the
goals. 

The LARS sample targeted star-forming galaxies (requiring H$\alpha$ equivalent 
widths above 100\,\r{A}) with FUV continuum luminosities similar to the 
$z\!\sim\!3$ LAEs and LBGs.  
The sample comprises over forty galaxies at redshifts $z\!=\!0.03-0.2$, 
observed with multiple HST 
broad-band and narrow-band filters, complemented with HST COS spectra, and
ground-based data including optical integral-field spectroscopy and
H\,{\sc i} radio interferometry. To obtain HST \lya\ images~\cite{Hayes13}, 
a dedicated method of broad-band filter subtraction was developed, due to the 
lack of existing appropriate narrow-band \lya\ filters.   

The UV-based selection produced a sample composed of irregular galaxies 
bearing signs of interactions, and disk/spiral galaxies. Their \lya\
shows a variety of properties, from absorption to bright emission, including
extended diffuse \lya\ halos that reach beyond the stellar or H$\alpha$
emission~\cite{Hayes13}. The high-resolution images comparing the 
stellar light, the optical gas emission and \lya\ thus show how \lya\  
travels from the productions sites (which should be identical to H$\alpha$)
to regions where it can easier escape. 
In complement, the UV and optical 
spectroscopy probe the effects of gas kinematics on the \lya\ 
escape~\cite{Rivera15,Herenz16}, while the radio 21cm data demonstrate the 
importance of the H\,{\sc i} mass~\cite{Pardy14}. 
Such detailed observations are by far not available for high-$z$ galaxies, and
therefore LARS, as its name indicates, provides a local reference sample 
for aiding the interpretation of high-$z$ observations.

\subsection{Lyman-break analogues}

As soon as the Lyman-break technique proved efficient for detecting galaxies
at high redshift (Section~\ref{sec:LBGs}), the question arose what type of 
galaxies were selected by this approach, especially in the situation where 
the FUV was the only available signal. 
The rest-frame UV properties were better mapped 
in high redshift than locally, 
and therefore it was impossible to draw analogies with nearby galaxies.  
Launch of the GALEX mission remedied this 
situation in the early 2000s and 
finally enabled the construction of statistical samples of low-$z$ UV galaxies.   

Lyman-break analogues (LBAs; formerly called UV-luminous galaxies, UVLGs)
were selected from the GALEX archive as galaxies at $z<0.2$ with 
FUV luminosities similar to 
those observed by the Lyman-break technique at high redshift \cite{Heckman05}.  
Such galaxies are rare in the local universe; 
the FUV selection resulted in $\sim\!70$ targets. 
The GALEX and SDSS images reveal that 
their morphologies range from compact systems with radii of 1\,kpc to large, 
late-type, 10\,kpc-size and $10^{11}M_\odot$ mass galaxies. 
The compact targets share many properties with the high-$z$ samples: 
they have stellar masses of $\sim10^{9.5-10.7}\,M_\odot$,
SFR\,$\sim5-25\,M_\odot\,\mathrm{yr}^{-1}$, and
metallicities 12+log(O/H)\,$\sim8.2-8.7$. 
Lyman-break galaxies at high $z$, on the other hand, 
can reach significantly higher SFRs 
(few hundred $M_\odot\,\mathrm{yr}^{-1}$) and they  
span a significantly larger metallicity range (12+log(O/H)\,$\sim7.7-8.8$).
Also, the local LBAs do not reach the maximum FUV luminosities of high-$z$
targets, the overlap between the FUV luminosities is achieved only 
close to the lower edge
of the high-$z$ FUV luminosity interval. 

Today, spectra and images at several wave bands 
are available for a subset of LBAs, including optical and UV data obtained 
with the HST, and X-ray data from Chandra. 
30\% of the large LBAs and 15\% of the compact
LBAs were found to be type-2 AGN \cite{Heckman05}   
(while type-1 AGN were removed from the sample upon selection).
Among the star-forming LBAs, a subset have been studied in more detail,  
revealing starburst character of their spectra, including strong optical 
emission lines, emission in the \lya\ line~\cite{Heckman11}, and 
in X-rays~\cite{Basu-Zych16,Brorby16}. Their X-ray emission is among the
brightest detected so far, with luminosity similar to high-redshift 
Lyman-break galaxies, and with resolved ULX sources or AGN candidates.
LBAs are also possible Lyman continuum leakers thanks to their fast stellar 
winds~\cite{Heckman11}. 
The Lyman continuum escape has indeed been confirmed for several LBAs
~\cite{Leitet13}.

\subsection{Green Peas and Luminous Compact Galaxies}
\label{sec:GPs}

\begin{figure}[bh]
\includegraphics[width=1.0\textwidth]{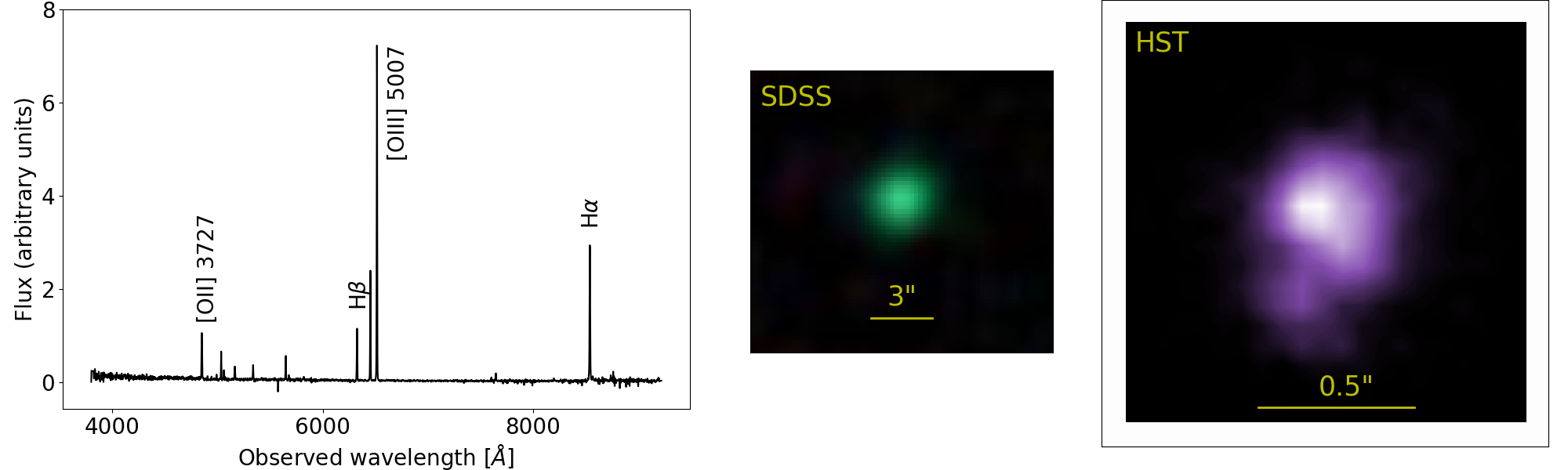}
\caption{
Green Pea optical SDSS spectrum (left); optical SDSS image -- 
unresolved (middle); and near-UV HST image -- obtained for Cosmic Spectrograph 
spectral acquisition
(right). All three panels are for the galaxy SDSS J092532.36+140313.0 at 
$z\sim0.3$. The diameter $0.5''$ corresponds to $\sim\!2$\,kpc.
Credit: SDSS and HST archives.}
\label{fig:GP}
\end{figure}	

Green Peas were identified in the SDSS by  
the citizen science Galaxy Zoo project
\cite{Cardamone09}. 
The targets that appear green and point-like, i.e. unresolved by the SDSS, 
turned out 
to be compact, highly star-forming galaxies at redshift $z\sim0.2.$
Their green colour is caused by bright optical emission lines. In particular,
the [O\,{\sc III}]\,$\lambda5007$ line reaches equivalent widths above
1000\,\AA\ (Figure~\ref{fig:GP}).  
It was shown later that the original eighty Green Peas are part of 
a more general population 
extending over a large interval of redshifts (at least $z\sim0.02 - 0.6$, 
probed by the SDSS), while they 
change colour accordingly with the emission line shift to other filters.
More than 800 such objects were identified in the SDSS and were 
named Luminous Compact Galaxies in \cite{Izotov11}.  

GP stellar masses are in the range $10^8-10^{10}\,M_\odot,$ their 
$\mathrm{SFR}\sim 1-60\,M_\odot$\,yr$^{-1}$, and oxygen abundances 
$12+\log(O/H) \sim 8.05-8.15$~\cite{Izotov11}. 
For those where UV data are available  
($\sim40$ targets \cite{Jaskot13,Henry15,Yang17,Orlitova18}), 
the Ly$\alpha$ line is observed in emission and  
many of the GPs would pass the selection criteria of high-$z$ 
Lyman-alpha emitters (see Section~\ref{sec:LAEs}).   
Their resolved HST images typically show a knotty structure 
(Figure~\ref{fig:GP}). Some of targets appear compact and structureless 
even with the HST resolution~\cite{Izotov18b}, which can however be a
question of data depth.

Do the Peas represent a new class or are they related to the other dwarf 
populations, such as the BCGs?
An evolutionary link is improbable, as the  
average GP masses and metallicities are larger (at an earlier cosmic time) 
than those of the average BCG.  
The continuity of the dwarf populations across redshifts is still being debated.
Nevertheless, both GPs and BCGs represent the rare types of 
nearby dwarf starburst galaxies that are more typical of high-$z$ universe, 
with low masses and metallicities, irregular morphologies, and low dust 
amounts.    
In contrast, there is a clear overlap between Green Peas and LBAs. 
While LBAs were UV-selected, the GPs were identified in the optical, 
but their follow-up UV observations show that they are conform with the 
LBA selection criteria.  

Green Peas (or Luminous Compact Galaxies) were among the first targets where 
a large escape of ionizing UV continuum was discovered 
\cite{Izotov16a,Izotov18b}, 
as we describe in Section~\ref{sec:LyC}. This strengthens their role of 
local laboratories for the reionization era galaxies ($z>6$).  
GPs are bright in X-rays~\cite{Svoboda19}, 
consistently with their high SFR and low metallicity. 
They are among the brightest star-forming galaxies known to date, exceeding the
empirical calibrations such as~\cite{Ranalli03,Brorby16}. 
This implies the presence of powerful X-ray sources, 
most likely the end products of massive stars, or potentially 
low-mass AGN~\cite{Kawamuro19}.
Nevertheless, this X-ray enhancement is not present in all GPs, some 
are X-ray sub-luminous~\cite{Svoboda19}.
This diversity in X-ray properties probably 
requires the dominance of extreme, short-lived  
sources such as ULXs.

\subsection{Dusty, luminous infrared galaxies (LIRGs)}
\label{sec:LIRG}

Galaxies with large amounts of dust are invisible or exceptionally 
faint in the UV and 
optical. Instead, their bolometric luminosities are dominated 
by the dust-reprocessed IR radiation.
Galaxies bright in the far-IR wavelengths were 
first reported in the 1960s, but dramatic progress was achieved with 
the IR satellites --
IRAS in the 1980s, Spitzer, Herschel and WISE after the year 2000
-- and the ground-based SCUBA array. 
The most luminous targets were classified (somewhat arbitrarily) 
into Luminous Infrared Galaxies (LIRGs) for luminosity 
$L_\mathrm{IR} \geq 10^{11} L_\odot$, 
and Ultra-luminous Infrared Galaxies (ULIRGs) for 
$L_\mathrm{IR} \geq 10^{12} L_\odot$.  
The dusty infrared galaxies are fascinating objects with colossal 
star-formation rates reaching 1000\,$M_\odot$\,yr$^{-1}$, which are among the 
largest known. 
Such a vigorous SF is necessarily 
short-lived because of the finite gas supply. The (U)LIRG may therefore 
represent an evolutionary phase, possibly coming after a violent
event such as a gas-rich galaxy merger. The merger would trigger the SF, which 
is subsequently responsible for the dust and for the obscured (U)LIRG. The  
obscured phase can possibly 
become an optically luminous AGN by blowing away the dust. This scheme
is still a matter of a lively debate, with uncertainties about its 
universality between different galaxy types, different redshifts and 
different environments. 
The literature about (U)LIRGs is vast, the puzzle of their origin and future
as well. We here describe the basic properties and we refer to the 
detailed reviews 
provided by Blain et al.~\cite{Blain02}, Lonsdale et al.~\cite{Lonsdale06}, and
Casey et al.~\cite{Casey14}, which contain 
a plethora of additional papers to read.

LIRGs and ULIRGs are massive galaxies, their morphologies include ellipticals, 
lenticulars, and spirals. 
Their dust temperatures vary between different targets, 
may vary with redshift, and multiple dust components are often present 
in the same galaxy (mostly ranging from 20 to 100\,K). 
It is interesting to note that \lya\ emission has been observed in some 
(U)LIRGs despite the large amounts of dust~\cite{Leitherer13}. 
Gas kinematics play a decisive role for the \lya\ escape there --  
high-velocity outflows (900\,km\,s$^{-1}$) were measured, using 
the UV absorption lines of metals. 
To account for the large IR luminosity, (U)LIRGs necessarily host 
a powerful UV source:  
an AGN, a starburst, or both. Optical studies have found 
AGN-like spectral lines in $>\!30\%$ ULIRGs, in parallel with  
evidence for a strong star formation in all targets.  
The incidence of AGN possibly increases with luminosity, some studies claim 
the AGN presence in more-or-less all ULIRGs. 
The difficulty of AGN confirmation lies in the ULIRG massive 
obscuration that makes even the
X-ray detection uncertain. ULIRGs were originally 
claimed sub-luminous in X-rays, but sensitive observations and 
polarized light observations discovered previously unknown active nuclei.     
Some of the ULIRGs host two or more AGN,  
and thus provide direct evidence of galaxy mergers. 
Deep optical and NIR imaging demonstrated that a vast majority of 
ULIRGs are located in interacting systems, ranging from widely separated 
galaxies to advanced mergers. Evidence of interaction between more than 
two galaxies led to the hypothesis that ULIRGs trace the previous presence 
of compact galaxy groups. Nevertheless, they never reside in rich environments
such as galaxy clusters. 

The causal relation between (U)LIRGs and quasars remains an open question 
that is being studied across redshifts by evaluating the clustering properties
of both populations as well as their internal structure.   
One hypothesis is that (U)LIRGs are obscured 
quasars seen edge-on, and can thus be part of the AGN unification scheme. 
A prevailing theory is the evolution of (U)LIRGs into unobscured 
quasars and then into elliptical 
galaxies~\cite{Lonsdale06,Alexander12}. 
ULIRGs represent only $<\!0.1\%$ galaxies in the local universe, 
but were much more common in the past, with a distribution  
peak at redshifts $z=1-3$. 
High redshift holds the key to understanding these galaxy populations 
that we will discuss more in Section~\ref{sec:SMGs}.

\section{Starburst galaxies at high redshift}
\label{sec:highz}

Starburst was a popular mode of star formation in the early universe,
unlike today where it is much rarer. 
High-$z$ starbursts formed a large fraction of today's stars and 
most probably provided the building blocks for galaxies 
as we know them today.

The knowledge attained to date about high-$z$ galaxies is less  
complete than that about the local universe. 
Until the 1990s, only AGN were observable at redshift $z\!>\!1$ 
(i.e. universe younger than 7\,Gyr). 
Since then, technological progress in telescopes and detectors 
has allowed probing galaxy populations out to $z\!\sim\!10$
(i.e. 500 Myr after the Big Bang).
Still, their low surface brightness and 
the small angular size only allow detection of their 
brightest features, mostly without angular resolution.
It is therefore convenient to design observational methods so as 
to target the prominent starburst signatures 
that we described in Section~\ref{sec:obs}. 
Essentially all galaxies detected so far at high $z$ 
are hence starbursts and   
the galaxies are classified by the detection method, instead of morphology or
other physical properties. 
The samples of brightest high-$z$ galaxies are complemented with 
lensed galaxies, 
which provide more detailed, deeper and spatially resolved information, 
reaching surface brightness
much fainter than would be detectable without the gravitational 
lens.

\subsection{Lyman-break galaxies} 
\label{sec:LBGs}

%
\begin{figure}[b]
\sidecaption
\centerline{\includegraphics[width=0.95\textwidth]{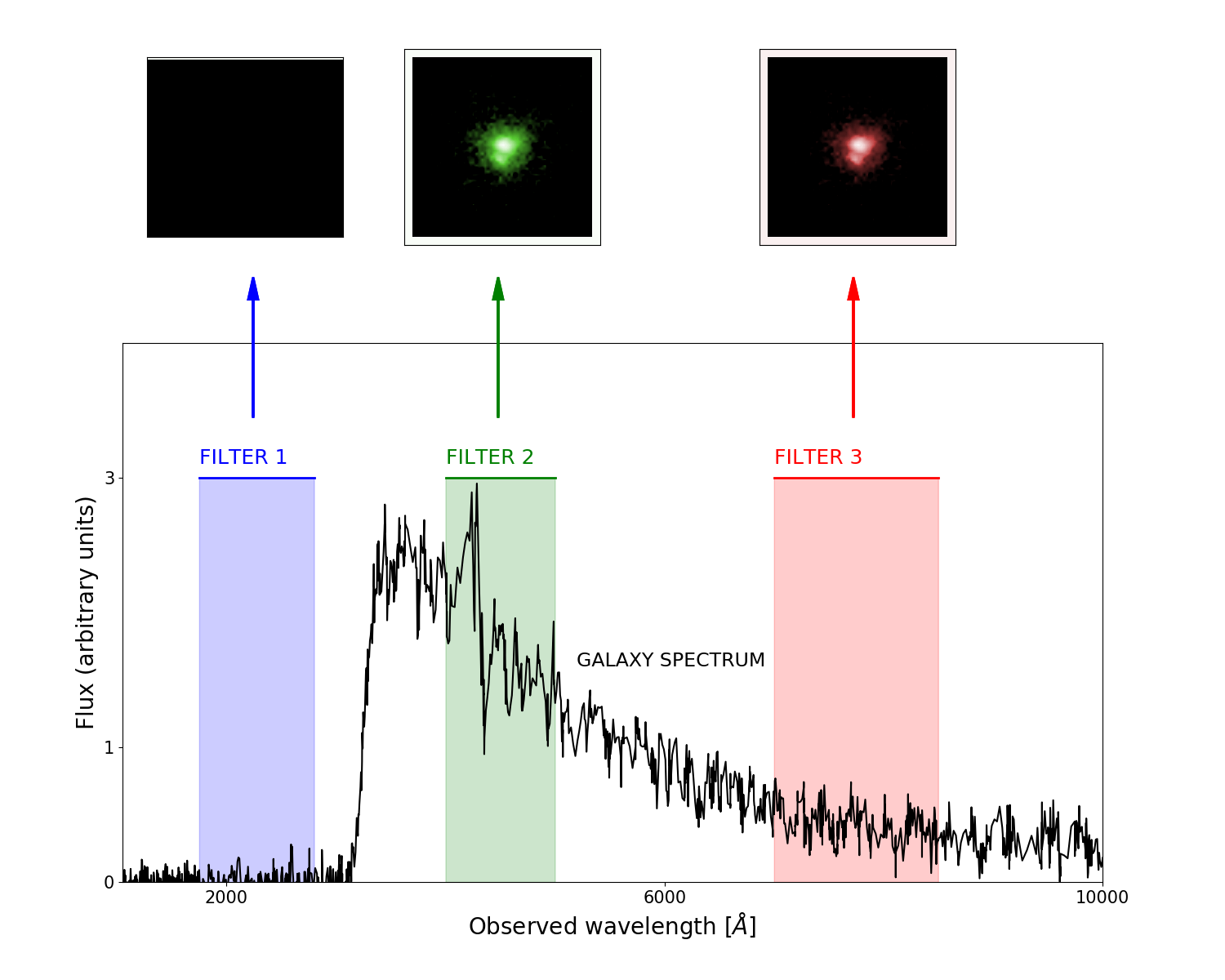}}
%
%
\caption{
LBG detection technique: (mock) galaxy is detected in a set of 
filters (here green and red), while no detection is achieved in the 
blue filter. The blue filter thus corresponds to wavelengths 
shorter than the Lyman break. 
A typical galaxy spectrum (rest-frame UV) is shown with the 
black line, here at redshift $z=2.5.$
}
\label{fig:LBG}       
\end{figure}

The Lyman break imaging method uses the fact that  
starbursts are UV-bright, with a sharp drop at the edge of the Lyman
sequence, i.e. $\lambda<912$\,\AA\  (Section~\ref{sec:UV}). 
The Lyman break is conveniently shifted to optical wavelengths at
$z\!>\!2.5,$ and to NIR at $z\!>\!7$, 
reachable by ground-based telescopes. 
The sky is imaged in a set of optical and NIR filters that sample the 
galaxy spectral energy distribution (SED). Cross-correlation 
of the sky maps in different filters will show 
galaxies that are detected in longer wavelengths 
and undetected in bluer filters. The disappearance of a galaxy from the 
filters corresponds to its Lyman break. The dropout filter provides 
its approximate redshift, 
with precision $\delta z\!\sim\!0.1-0.2$ 
(Figure~\ref{fig:LBG}). 
Objects detected by this technique are referred to as 
Lyman-break galaxies (LBGs) or dropout galaxies.  
The advantage of the method is the use of broad-band filters,
and thus an efficient retrieval of large galaxy samples.  
The precise redshifts are then secured with 
follow-up spectroscopy that requires significantly longer exposure times.  
As we approach the reionization era, the \lya\ forest 
becomes so optically thick that it removes essentially all the flux 
at $\lambda<1216$\,\r{A} (Section~\ref{sec:reion}), 
and therefore the Lyman break technique 
actually becomes the Gunn-Peterson trough technique, detecting the sharp drop
blueward of \lya.

After the pioneering steps in the LBG technique, breakthrough discoveries  
were done by 
Steidel et al. \cite{Steidel96} at $z\sim3$. They started
with 4-m telescopes such as the William Herschel Telescope in 1993 and 
then switched to the 10-m Keck Telescope and to the Hubble Space Telescope 
(Hubble Deep Field survey). 
Nowadays, the known LBG samples count tens of thousands of targets,  
ninety percent of which are situated in the 
redshift interval $z\!\sim\!2.5-3.5$. 
Observations at $z\!>\!5$ 
became possible later, with the development of more sensitive instruments, 
which was especially true for 
$z\!>\!7$, where the Lyman break shifts to the NIR. 
The Hubble Deep Field, Hubble
Ultra-Deep Field, and the GOODS survey have been prolific at identifying
LBG populations out to redshifts $z\!=\!7-10$~\cite{Vulcani17}. In addition,  
galaxies detected through the \lya\ line (Section~\ref{sec:Lya})
may form a subset of the LBGs and provide thus complementary information. 

The typical LBGs have stellar masses ranging from $10^9$ to 
$10^{11}M_\odot$, and metallicities from 10\% solar to solar. 
More detailed information is available for a few LBGs 
that are in fortuitous 
alignment with a foreground galaxy and are thus gravitationally lensed:
e.g. the Cosmic Horseshoe~\cite{Quider09}, the Cosmic Eye~\cite{Quider10}, 
the 8 o'clock arc~\cite{Dessauges10}, or the Sunburst Arc~\cite{Rivera17}. 
The lensing provides clues about their geometry, ISM composition 
and kinematics, about 
stellar populations, and about the escape of ionizing radiation.  
Most of the lensed LBGs seem to be rotating disk galaxies with multiple 
giant star-forming H\,{\sc ii} regions that have a
SFR density two orders higher
than spiral galaxies seen in the local universe. 
The comoving number density of LBGs is similar to present-day  
bright early-type galaxies, and therefore an evolutionary link between the two
is possible. 
However, whether their evolution proceeded through major mergers, 
or they are cores of present-day galaxies which evolved passively without major
events, still needs to be clarified. To this end, deep multi-wavelength 
spectroscopic follow-ups are necessary (and challenging) 
in order to constrain the morphologies, precise masses,   
velocity fields, duration of the star-forming activity, and evolution of 
LBGs across redshifts. Useful review papers were provided by 
Giavalisco~\cite{Giavalisco02} and Dunlop~\cite{Dunlop12}.

\subsection{Lyman-alpha emitters (LAEs)}
\label{sec:LAEs}
Lyman-alpha emitting galaxies were important constituents of the 
high-$z$ universe, as they represent the majority of the known 
star-forming galaxies at redshifts close to reionization.   
Using \lya\ as a detection tool was proposed as early as the 1960s 
\cite{Partridge67}. 
However, the first \lya\ detections were not available 
until the 1980s~\cite{Djorgovski85} when sizes and sensitivities of 
telescopes and detectors became sufficient.          
The \lya\ fluxes were several factors lower than expected, as we explained 
in Section~\ref{sec:Lya}. Still today, predicting the \lya\ luminosity for an
individual galaxy is not straightforward due to the multi-parameter 
nature of the \lya\ escape. 
The largest ground-based telescopes and space telescopes (HST and soon JWST) 
are used for observing high-$z$ LAEs today. 
The asymmetric \lya\ spectral profile (Section~\ref{sec:Lya}, 
Figure~\ref{fig:Lya}) 
presents an advantage for high-$z$ surveys, where it
allows a unique identification of the line and hence of the galaxy redshift.  
Recently, LAE surveys have gained another dimension 
thanks to the use of 
integral-field spectrographs such as MUSE at the 
ESO VLT~\cite{Wisotzki16}. Simultaneous observation of spectra  
from different parts of the galaxy allows testing the theories of \lya\ 
transfer from the production sites all the way to the outskirts.

Narrow-band \lya\ imaging surveys pick 
galaxies where \lya\ is produced in large quantities and is only weakly 
attenuated. This requires a powerful starburst activity together with 
low amounts of dust and neutral hydrogen, as well as the presence 
of outflows that shift \lya\ out of resonance. 
These conditions are met in low-mass galaxies 
(typically $10^8-10^{10}$\,M$_\odot$), but not only.  
The low stellar mass facilitates the action of stellar feedback that can 
more easily outweigh the gravitational potential. 
However, the conditions that allow \lya\ to escape span a certain range 
(Section~\ref{sec:Lya}), 
and therefore the possible LAE masses and dust amounts can be larger  
in special cases.  
The escape of \lya\ photons may also be non-isotropic due to the disk galaxy 
morphology~\cite{Verhamme12}, due to gas clumpiness or 
preferential directions of escape~\cite{Behrens14,Verhamme15}. 
To improve our understanding of what galaxy population is selected by 
\lya, several studies have searched for the differences 
between LAEs and the continuum-selected LBG or the emission-line-selected 
(in the sense of rest-frame optical) galaxies on the same redshifts. 
No statistical differences were found between LAEs and optical emission-line 
galaxies in the morphology, inclination, sSFR, UV slope or the distance between
neighbours~\cite{Hagen16}, suggesting that the \lya\ escape is driven by the 
detailed local physics rather than by global properties. 
Some LBGs are bright in \lya, therefore the LBG and LAE 
populations at least partially overlap. 
The mean LAE mass is usually found to be 
one or two orders of magnitude lower than the mean LBG mass.
However, there are studies that claim no differences between the two 
populations (discussed in~\cite{Hagen16}). 
On the other hand, clustering properties of LAEs and LBGs seem to  
differ, suggesting their different evolutionary 
paths~\cite{Ouchi09b,Dunlop12}. 

The fraction of \lya\ photons that escape from galaxies 
evolves with redshift~\cite{Hayes11}: 
the average escape fraction 
increases from $z\!=\!0$ to $z\!=\!6$ and declines at $z\!>\!6.5$. 
The increase  
correlates with a decreasing dust amount in the cosmos,  
suggesting that dust could be  
the main driver of the {\em average\/} \lya\ escape 
(modulated by other parameters~\cite{Atek14}).
The escape fraction drop at $z\!>\!6.5$ can be attributed to the progressively 
more neutral IGM   
i.e. an effect of the environment rather than the galaxies themselves. 
The drop may also be an intrinsic feature of the reionization era galaxies, 
either
due to a large fraction of static H\,{\sc i} (as in the local IZw18) 
or the opposite, the Lyman continuum escape. 
At redshift $z\!>\!6$, not only the escape fraction, but also 
the number of observed LAEs decreases, 
either due their intrinsically lower numbers 
or due to the neutral IGM. 
LAEs are currently observed out to redshifts 
$z\!>\!10$, and play thus an important role as cosmological 
tools~\cite{Ouchi09b,Sobral15,Zitrin15,Oesch16}.
Clustering properties map the dark matter structures and the evolution of 
LAEs through redshifts to the present day. The clustering 
suggests that they were the building 
blocks of the present-day MW-type galaxies~\cite{Guaita10}.
The LAE clustering 
in the reionization era 
probably also maps the patchy IGM ionization. 
The importance of LAEs is reflected in the objectives of 
the new generation of telescopes, such as the JWST or the ELTs, with the 
ambition to use \lya\ to detect the first galaxies in the universe.

\subsection{Sub-millimeter galaxies}
\label{sec:SMGs}

High-redshift counterparts of local ULIRGs, i.e. dust-obscured, 
massive starbursts (Section~\ref{sec:LIRG}), are known under the name 
Sub-millimeter galaxies (SMGs).
Their detection has been effectively possible since the  
operation of the SCUBA bolometer array 
on Mauna Kea in the 1990s (see review~\cite{Blain02}). 
Nowadays, several sensitive mm/sub-mm telescopes 
are available, including IRAM (Spain), NOEMA (France), or ALMA (Chile). 
To facilitate the orientation in the vast literature on the subject, 
we refer preferentially to review papers in this text.  

Typically scaled-up versions of ULIRGs, the SMGs are the most intense 
SF sites and the most bolometrically luminous galaxies in the 
universe ($L_\mathrm{bol}\sim10^{13} L_\odot$). 
Their molecular gas masses reach $10^{11}M_\odot$ and they fuel 
star-formation rates exceeding 1000\,$M_\odot$\,yr$^{-1}$.
ULIRGs were once thought to be evolved versions of SMGs. 
However, today's view is that SMGs were more massive 
and cooler than ULIRGs and thus 
any evolutionary link between them is unlikely.
While the number density of nearby ULIRGs is one per four square degrees, 
there were hundreds per one square degree at $z\!>\!1$~\cite{Lonsdale06}.  
Cosmologically, SMGs should be the tracers of massive dark matter halos 
and their evolution across cosmic history. Recent instrumental advances 
(especially ALMA) have permitted the first studies of SMG 
clustering 
properties (reviewed in~\cite{Casey14}), which will allow testing the theories
of the SMG origin and their fate. 
High-$z$ SMGs were predicted to reside
in rich environments where they would accumulate their mass. 
According to today's knowledge, SMGs mostly appear to 
live in relative isolation (contrary to the predictions),  
even though they bear signs of galaxy mergers. Only the 
most luminous SMGs have been identified to reside in potential proto-clusters
of galaxies~\cite{Casey14}.

SMGs show a surprisingly diverse range of optical properties, 
from undetected to bright sources. 
The optical spectra provide evidence for both starburst and AGN,
residing in metal-rich gas.  
The bolometric luminosity is typically dominated by star formation. 
The lower limit for AGN incidence in SMGs is 30\%, though in reality is 
probably 
significantly larger. The incidence seems to increase with luminosity and 
therefore with redshift. Intriguing results have been provided by 
radio observations showing
that $\sim\!70\%$ of SMGs are spatially extended on $\sim\!10$\,kpc scales,  
i.e. dramatically larger than the dust emission that tends to be 
concentrated in $<\!1$\,kpc. Several interpretations of the 
radio emission exist, ranging from 
large-scale SF to jet-like structures similar to those in radio AGN
~\cite{Lonsdale06}.
SMGs are also known as sites of gamma-ray bursts (GRBs), ignited at 
the death of massive stars. 

The present-day mainstream hypothesis is that SMGs evolved into  
quasars and eventually into early-type galaxies (nicely depicted 
in~\cite{Alexander12}). The SMG masses and SFRs indeed make     
them the ideal progenitors of elliptical galaxies. The existence of 
large numbers of massive galaxies such as SMGs out to high redshift 
represents a 
challenge to the simple hierarchical galaxy formation paradigm that 
predicted slow mass buildup through small galaxy mergers. 
On the other hand, it is not in contradiction with dark matter 
halo growth, which allows for rapid baryon accumulation in very massive
halos~\cite{Lonsdale06}.   
The SMGs, their place in galaxy evolution, their progenitors and 
their successors 
thus play an important role in answering fundamental astrophysical questions 
such as the 
buildup of stellar mass in the universe, the origin of present-day galaxies, 
and the growth of supermassive black holes.

\section{Role of starburst galaxies in reionizing the universe}
\label{sec:cosmic}

\subsection{Cosmic Reionization} 
\label{sec:reion}

The present-day intergalactic medium is essentially fully ionized, but
it was not so during all of the cosmic history. 
After the early evolutionary stages where 
matter and radiation were constantly interacting, 
the universe became cool enough for radiation to 
split from matter at $z\sim1000$ 
(i.e. $\sim300\,000$ years after the Big Bang). 
The radiation then freely travelled through the universe without 
being absorbed by atoms.  
This was the start of Dark Ages, where the universe was neutral 
 and without sources of radiation (apart from the H\,{\sc i} line at 21\,cm).  
The first stars are supposed to have formed at $z\sim30$. 
Current observations have not yet reached such distant redshifts to capture 
their light. Nevertheless, an exceptional result was recently achieved: 
an absorption feature was observed in the H\,{\sc i}~21\,cm line profile, 
consistent with theoretical predictions of the effect of the first 
stars~\cite{Bowman18}. The observation thus proved that stars had been 
formed by $z\sim20$.
The phenomenon will further be explored by upcoming surveys
such as LOFAR, HERA or SKA. 

The birth of the first sources of light caused a phase change 
in the IGM:  
the IGM progressively evolved from neutral to a fully ionized state. 
The era between the appearance of the first stars and the 
redshift $z\sim6$ where the IGM became essentially completely 
ionized is called the Cosmic 
Reionization. The end of reionization is well demonstrated by the Gunn-Peterson 
effect in quasars~\cite{Gunn65}: the FUV light of a distant quasar (or another
bright source) encounters hydrogen clouds along the line of sight and 
the H\,{\sc i} scatters away the light that gets in resonance with \lya\ at the
cloud's redshift. At different redshifts, different parts of the spectrum 
``become'' \lya, and, as a result, the original spectrum adopts 
numerous absorption features, known as the ``\lya\ forest''.
At $z\!>\!6$, the discrete features turn into 
a continuous Gunn-Peterson trough that removes all the 
FUV at wavelengths shorter than the quasar's rest-frame 
\lya~\cite{Becker15}.  Additional constraints for reionization 
come from the cosmic microwave background radiation 
measured with the WMAP and Planck satellites \cite{Robertson13,Robertson15}.

\subsection{Sources of reionization -- starburst galaxies?}
\label{sec:LyC}

Details of the cosmic reionization process depend on the density 
and distribution of gas, and, most importantly, on the available ionization 
sources, their density, distribution, evolution and energy spectrum 
(see review~\cite{Robertson10}). 
The first galaxies 
provided ionizing FUV radiation from hot, massive stars, and are thus
natural candidates for reionization.  
Furthermore, quasars set in at some point, 
providing copious amounts of energetic radiation. Additional sources such 
as X-ray binaries or cosmic rays are being considered as possible 
ionization contributors (see references in \cite{Svoboda19}). 

The distribution of the first galaxies was associated with dark matter 
structures, 
which are possible to model using cosmological simulations.   
Their ionizing photon production rate 
depends on the nature of the first stars and on the star 
formation rate density,  
for both of which we need an observational input in order to achieve 
realistic predictions~\cite{Robertson10}.
However, yet another parameter controls the ionizing 
radiation (Lyman continuum) flux to the IGM: its
escape fraction, reflecting 
the galaxy's ISM structure (total H\,{\sc i} amount, dust, asymmetry, 
clumpiness). 
Numerical simulations predict that despite a 
strong variability of the escape fraction in time and in   
angular direction, 
an average value of $\sim\!20\%$ per galaxy is necessary
to achieve the reionization solely by star-forming 
galaxies~\cite{Paardekooper15}. 
On the observational side, 
direct detection of the Lyman continuum from the reionization era galaxies is 
unfeasible: any ionizing photon that escapes from the galaxy will be absorbed
by the neutral IGM and will not reach our telescopes. 
Only indirect signatures can be explored at the reionization era, 
while direct studies of the Lyman continuum leakage are restricted to lower
redshifts, in practice $z\!\lesssim\!3$, where the IGM H\,{\sc i} amount is 
sufficiently low. 

Searches for escaping Lyman continuum at any redshift 
were fruitless during two decades of observational efforts, 
except for a few sources with low leakage fractions 
\cite{Leitet13,Borthakur14}. 
The situation has dramatically changed in the past few years. 
In 2016-2019, Lyman continuum was spectroscopically confirmed in 
approximately forty $z\!\sim\!3$ 
galaxies~\cite{Vanzella16,Shapley16,Bian17,Steidel18,Fletcher19}. 
In parallel, a significant leakage 
was detected with the HST in eleven starburst galaxies of Green-Pea type 
at $z\!\sim\!0.3$ (summarized in \cite{Izotov18b}). 
The Lyman continuum escape fraction reaches 70\% in some of these
targets, while their mean is $\sim\!20\%$. Interestingly, this percentage 
is consistent with the numerical predictions mentioned above. 
Despite their too low redshift, the leaking galaxies support 
the possibility of cosmic reionization by their high-redshift analogues.   

The low-$z$ samples serve as laboratories for understanding  
the physical conditions that lead to a low optical depth at  
Lyman continuum wavelengths.
From the current observations, we may speculate about the role of 
the galaxy compactness and low mass, which favour the escape of ISM 
gas from a shallow potential well and thus decrease the H\,{\sc i} column.  
Second, low metallicity 
modifies stellar evolution, leading to a larger production 
of ionizing photons. Third, feedback from 
starburst-related processes (stellar winds, energetic photons and 
jets/outflows from compact binary systems) deposits radiative and 
mechanical energy into the ISM, which results in gas ionization
and/or its removal along certain paths.    
Indirect indicators are being explored 
that would help us preselecting galaxies with Lyman continuum escape: 
various works probe the ISM ionization~\cite{Nakajima14,Stasinska15}, 
ISM velocities~\cite{Heckman11}, saturation of the ISM spectral 
lines~\cite{Chisholm18},  
X-ray emission~\cite{Svoboda19}, and the \lya\ line 
profile~\cite{Verhamme15,Verhamme17}.
So far, the \lya\ line
was proven, both theoretically and observationally, to be the best 
indicator of the Lyman continuum escape, 
thanks to its sensitivity to the ISM 
properties~\cite{Verhamme15,Verhamme17,Izotov18b}.

A natural question is whether quasars, 
efficient producers of ionizing photons, could be the true sources 
of cosmic reionization instead of the star-forming galaxies. 
The question remains open so far.
The first quasars formed in the reionization era, 
but it seems that they were not abundant enough 
at $z>6$ to dominate the reionization. 
However, with the scarcity of data, any new detection may change the 
picture~\cite{Giallongo15,Madau15,Mitra18}. 
Nevertheless, the Lyman continuum of starburst galaxies 
appears to be the most viable mechanism for reionization, winning over quasars 
or any other sources.

\section{Conclusions and future prospects}
\label{sec:concl}

It has been estimated in the literature that equal amounts of stars 
in the universe were formed in the following four types of environments: 
optically visible regions, dust-enshrouded regions of optical galaxies, 
heavily obscured galaxies with $L_\mathrm{IR} <10^{12} L_\odot,$ and 
ULIRGs with $L_\mathrm{IR}>10^{12} L_\odot$~\cite{Trentham04}. 
Dwarf galaxies dominated the star formation among the optically visible ones,
while massive starburst galaxies are dusty and IR-bright.  
We have reviewed the different types of starburst galaxies both at low and
high redshift. 
Without being complete, this overview illustrated the various 
 possible manifestations of vigorous star formation,      
the corresponding detection techniques, the properties of various 
starburst galaxy populations, and their role in cosmic evolution.
We demonstrated the search for
analogies between low and high-redshift galaxies.
Analogies allow complementary views on star formation and galaxy evolution,    
and on the origin and the future of starburst galaxy populations. 
In light of the analogies with high redshift, local galaxies 
have seen an increased interest in their classification by parameters 
such as the SFR, metallicity, FUV luminosity or compactness.       
While the overlap between high-$z$ and low-$z$ galaxy parameters cannot 
be perfect, the nearby galaxies serve as local laboratories that provide  
essential clues for the interpretation of processes in the 
distant universe where the details are unattainable.

Research on starburst galaxies has good prospects for future, and 
in particular galaxies at the cosmic dawn
are among the main drivers for the new generation of astronomical 
instruments. 
MUSE, a sensitive optical integral-field spectrograph is already 
operating at ESO VLT.  
Its ability to simultaneously obtain hundreds of spectra 
per galaxy 
combines the advantages of classical imaging and spectroscopy.
In combination with adaptive optics, MUSE reaches an angular resolution 
comparable to that of the HST and it can outperform the HST in detecting
distant galaxies \cite{Bacon15,Bacon17}.   
Spatially resolved spectra of LAEs observed at $z\!\in\!(3,6)$ with
MUSE  have brought unprecedented details 
on the structure of these distant 
starbursts and resolved individually their diffuse \lya\  
halos~\cite{Wisotzki16,Wisotzki18}.
ALMA, a recently built international facility in Chile, 
is an interferometer composed of 66 antennas 
observing at mm and sub-mm wavelengths. 
It is well suited for observing dusty starbursts and SMGs at 
$z>2$, their distribution across redshifts, and possibly their evolution into 
massive early-type galaxies~\cite{Karim13,Dunlop17}. 
ALMA is also efficient at mapping the distribution of dust in low-mass 
predecessors of MW-type galaxies,  
such as the UV-selected LBGs and LAEs \cite{Riechers14} 
or the Chandra-detected X-ray galaxies \cite{Karim13}. 
ALMA has a sufficient sensitivity to resolve high-$z$ infrared atomic and 
molecular lines that 
probe galaxy metallicities, ionizations and kinematics, and determine 
their precise redshifts~\cite{Vieira13,Hashimoto18}. 
In the local universe, ALMA is mapping hundreds of thousands of  
stellar nurseries in nearby galaxies with the goal to shed light on the
star-formation mechanisms and the conditions under which the 
formation proceeds by intense bursts. 

In the years to come, several major instruments will become active.   
HARMONI is a visible/NIR integral-field spectrograph 
in construction for the  E-ELT's first light in the mid-2020s.  
Assisted with adaptive optics, the HARMONI angular 
resolution will be comparable to that achieved with space-borne instruments. 
The optical/NIR wavelengths will detect rest-frame UV (\lya, continuum, lines) 
in the reionization era.  One of the goals is the 
detection of Population-III stars, i.e. primordial stars with no heavy 
elements, possibly reaching hundreds of solar masses. 
Their detection through the ionized helium lines has been claimed possible  
with HARMONI out to $z\!>\!10$. HARMONI will thus play an important 
role in  probing the high-mass end of the IMF.
In the local universe, HARMONI will resolve 
individual stars in nearby galaxies and supernovae out to $z\sim3-4$,
testing our models of star-formation and galaxy assembly histories.     
HARMONI already has a dedicated simulation tool to predict the 
future observational outcomes, using mock data from  
cosmological simulations \cite{Augustin19}. The tool 
may also be an essential preparatory step for observers once the instrument 
is operational. 

Space missions play an ever larger role in cosmic studies.  
The JWST will operate in the IR domain and will detect the rest-frame UV
of reionization era galaxies.  
The telescope will possess the 
first space-borne integral-field spectrograph, NIRSPEC.
Compared to the ground-based instruments, NIRSPEC will have much more severe
limitations on the number of spectra that it can cover simultaneously
(100 as opposed to 31\,000\ in HARMONI). On the other hand, it will 
have the advantages of the space mission, unaffected by the Earth's atmosphere.
Among other future missions, ESA's EUCLID will map 
the large cosmic structure and its evolution. 
It will produce deep optical 
and NIR images and spectra of galaxies out to $z\!\sim\!2$ across 
approximately half of the sky. 
The unprecedented extent of this catalogue will provide the basis for 
statistical studies mapping the galaxy assembly in the past ten billion years.
In X-rays, the newly planned mission ATHENA
 will have a high sensitivity and high angular resolution.  
It will complete the census of evolved binary sources in starburst 
regions down to faint X-ray fluxes. It will resolve the spatial 
structure of X-ray emitting regions in galaxies and will constrain the 
dominant sources, discriminating 
between X-ray binaries, AGN, hot gas and other.  
Eventually, ATHENA will open the X-ray window for high-redshift galaxies, 
too faint to be reachable individually by current instruments. 

Complementary to direct detections of starburst signatures, 
radio telescopes such as SKA, HERA or LOFAR will map the atomic hydrogen  
distribution in the universe   
at its different epochs, including the reionization era. 
Using arrays of hundreds of radio antennas, these interferometers 
will resolve how the
neutral IGM structures evolved, in what conditions formed  
the first stars and galaxies and how they 
continue to form today.

\bigskip

Astronomy is experiencing a golden age and the quest for understanding how 
stars formed and
how galaxies assembled through the cosmic history is one of its flagships. 
The existing and upcoming facilities, using multiple spectral windows,
take us progressively to the very first structures that formed in the universe. 
Supported by numerical simulations and by detailed, 
resolved local studies, they will complete our picture of the conditions 
for star formation at different epochs, the origin of galaxies as we know 
them today, the existence of extreme objects, and the mechanisms driving galaxy 
evolution.

\begin{acknowledgement}
The author is grateful to the referee for their thoughtful 
comments that made this review more complete. 
The author was supported by the Czech Science Foundation project 17-06217Y 
while working on the manuscript. 
\end{acknowledgement}
%


\bibliographystyle{spphys}
\bibliography{references}

\end{document}